\documentclass{jetpl}
\usepackage {cite}
\twocolumn

\lat

\title{Kohn-Luttinger superconductivity in monolayer
and bilayer semimetals with the Dirac spectrum}

\rtitle{Kohn-Luttinger superconductivity in monolayer and bilayer
semimetals with the Dirac spectrum}

\sodtitle{Kohn-Luttinger superconductivity in monolayer and
bilayer semimetals with the Dirac spectrum}

\author{M.\,Yu.\,Kagan$^{a,b}$,
V.\,A. Mitskan$^{c,d}$, M.\,M. Korovushkin$^{c}$}

\rauthor{M.\,Yu.\,Kagan, V.\,A. Mitskan, M.\,M. Korovushkin}

\sodauthor{}

\address{$^a$Kapitza Institute of Physical Problems, Russian Academy of Sciences, Moscow, 119334 Russia\\
$^b$National Research University Higher School of Economics, Moscow, 109028 Russia\\
$^c$Kirensky Institute of Physics, Siberian Branch, Russian Academy of Sciences, Krasnoyarsk, 660036 Russia\\
$^d$Reshetnev Siberian State Aerospace University, Krasnoyarsk, 660014 Russia\\
}

\abstract{The effect of Coulomb interaction in an ensemble of Dirac fermions on the formation of superconducting pairing in monolayer and bilayer doped graphene is studied using the Kohn-Luttinger mechanism disregarding the Van der Waals potential of the substrate and impurities. The electronic structure of
graphene is described using the Shubin-Vonsovsky model taking into account the intratomic, interatomic,
and interlayer (in the case of bilayer graphene) Coulomb interactions between electrons. The Cooper instability is determined by solving the Bethe-Saltpeter integral equation. The renormalized scattering amplitude
is obtained with allowance for the Kohn-Luttinger polarization contributions up to the second order of perturbation theory in the Coulomb interaction. It plays the role of effective interaction in the Bethe-Salpeter integral equation. It is shown that the allowance for the Kohn-Luttinger renormalizations as well as intersite
Coulomb interaction noticeably affects the competition between the superconducting phases with the $f-$wave
and $d + id-$wave symmetries of the order parameter. It is demonstrated that the superconducting transition
temperature for an idealized graphene bilayer with significant interlayer Coulomb interaction between electrons is noticeably higher than in the monolayer case.
}

\begin{document}

\maketitle

\section{INTRODUCTION}

Graphene is of considerable interest for fundamental physics
and for applications due to its peculiar
transport, pseudorelativistic, and quantum-electrodynamic
properties ~\cite{Lozovik08,Castro09,Kotov12}. This combination of graphene
properties is primarily determined by its unique gapless
energy structure consisting of cone-shaped valence and
conduction bands contacting at the corners of the first
Brillouin zone (Dirac points)~\cite{Wallace47}. It has been
established that electrons propagating in graphene near
Dirac points resemble massless fermions with linear
dispersion~\cite{Novoselov05} and are characterized by the minimal
conductivity for a zero charge carrier concentration~\cite{Novoselov05,Tan07},
a high mobility~\cite{Morozov08,Bolotin08,Garcia08}, Klein tunneling~\cite{Geim06,Young09},
oscillating motion (Zitterbewegung)~\cite{Katsnelson06,Rusin09}, universal
absorption of light~\cite{Nair08}, and many other properties
having no analogs in other physical systems.

When in contact with superconductors, graphene
exhibits exotic superconducting properties~\cite{Munoz12}. In
spite of the fact that the evolution of the Cooper
instability in graphene itself has not yet been confirmed,
experimental evidence~\cite{Heersche07,Shailos07,Du08,Ojeda09,Kanda10,Tomori10}
 that graphene in contact
with conventional superconductors exhibits superconducting properties have been obtained. The fact
that short graphene samples placed between superconducting
contacts can be used to construct Josephson
junctions indicates that Cooper pairs can propagate
coherently in graphene. In this connection, it would be
interesting to find out whether it is possible to modify
graphene structurally or chemically to convert it into a
magnet~\cite{Peres05} or even into a real superconductor.

It is known theoretically that a model with conic dispersion
requires the minimal intensity of the pairing interaction for the
development of the Cooper instability~\cite{Marino06}. In this
connection, there have been several attempts to analyze
theoretically possible achievement of the superconducting state in
doped monolayer, as well as bilayer, graphene. The role of
topological defects in achieving Cooper pairing in a graphene
monolayer was studied in~\cite{Gonzalez01}. In~\cite{Uchoa07}, the
phase diagram for spin-singlet superconductivity in a monolayer
was constructed by Uchoa and Castro Neto in the mean field
approximation, and the plasmon mechanism of superconductivity
leading to low superconducting transition temperatures in the
$s$-wave channel was studied for realistic electron concentration
values. The possibility of inducing superconductivity in a
graphene monolayer due to electron correlations was investigated
in~\cite{Black07,Honerkamp08}.

The situation in which the Fermi level is near one of
the van Hove singularities in the density of states of a
graphene monolayer was considered in~\cite{Gonzalez08}. It is well
known that these singularities can enhance magnetic
and superconducting fluctuations~\cite{Markiewicz97}. In accordance
with the scenario described in~\cite{Markiewicz97}, the Cooper instability
occurs due to strong anisotropy of the Fermi contour
for van Hove filling $n_{vH}$, which in fact originates from the
Kohn-Luttinger mechanism~\cite{Kohn65} proposed in 1965 and
assuming the occurrence of superconducting pairing in
systems with purely repulsive interaction~\cite{Fay68,Kagan88,Baranov92}. It was
noted in~\cite{Gonzalez08} that this mechanism can occur in graphene
because the electron-electron scattering becomes
strongly anisotropic; for this reason, a channel with
attraction can be formed when there is a projection onto
harmonics with a nontrivial angular dependence on the
Fermi surface. According to the result obtained in~\cite{Gonzalez08},
such the Cooper instability in an idealized graphene
monolayer evolves predominantly in the $d-$wave channel
and can be responsible for superconducting transition
temperatures up to $T_c\sim10\,K$ depending on the proximity
of the chemical potential level to the van Hove singularity.
An analogous conclusion was drawn in~\cite{Nandkishore12a}, where
the Kohn-Luttinger superconductivity in the vicinity of
the van Hove singularity in the graphene monolayer was
studied by the renormalization group method.

The possibility of the competition and coexistence of the
Pomeranchuk instability and the Kohn-Luttinger superconducting
instability in a graphene monolayer was considered
in~\cite{Valenzuela08}. In~\cite{McChesney10}, it was found in
experiments with a strongly doped monolayer using angle-resolved
photoemission spectroscopy (ARPES) that multiparticle interactions
substantially deform the Fermi surface, leading to an extended van
Hove singularity at point $M$ of the hexagonal Brillouin zone. The
features of the ground state were investigated theoretically, and
the competition between the ferromagnetic and superconducting
instabilities was analyzed. It was shown that in this competition,
the tendency to superconductivity due to strong modulation of the
effective interaction along the Fermi contour (i.e., due to
electron-electron interactions alone) prevails. The
superconducting instability evolves predominantly in the $f-$wave
channel in this case~\cite{McChesney10}. The competition between
the superconducting phase and the spin density wave phase at van
Hove filling and near it in the monolayer was analyzed
in~\cite{Kiesel12} by the functionalization renormalization group
method. It was found that for the band structure parameters and
the Coulomb interactions obtained by ab initio calculations for
graphene and graphite monolayers~\cite{Wehling11},
superconductivity with the $d + id-$wave type of symmetry of the
order parameter prevails in a large domain near the van Hove
singularity, and a change in the related parameters may lead to a
transition to the phase of the spin density wave. According
to~\cite{Kiesel12}, far away from the van Hove singularity, the
long-range Coulomb interactions change the form of the $d +
id-$wave function of a Cooper pair and can facilitate
superconductivity with the $f-$wave symmetry of the order
parameter.

In accordance with the results obtained in~\cite{Gonzalez13}, in
the case of bilayer graphene, ferromagnetic instability in the
vicinity of the van Hove singularities dominates over the
Kohn-Luttinger superconductivity. However, the Coulomb interaction
screening function in the bilayer was calculated earlier
in~\cite{Hwang08} in the random phase approximation (RPA) in the
doped and undoped regimes. It was found that the static
polarization operator of the doped bilayer contains the Kohn
anomaly much larger than in the case of a monolayer or a 2D
electron gas. It is well known that the singular part of the
polarization operator or the Kohn
anomaly~\cite{Migdal58,Kohn59,Friedel54} facilitates effective
attraction between two particles, ensuring a contribution that
always exceeds the repulsive contribution associated with the
regular part of the polarization operator for the orbital angular
momenta $l\neq0$ of the pair~\cite{Kohn65}. Thus, we can expect
that the superconducting transition temperature $T_c$ in the
idealized bilayer may exceed the corresponding value for the
idealized monolayer.

In addition, it was shown in earlier
publications~\cite{Kagan91,KaganValkov11} that the value of $T_c$
can be increased via the Kohn- Luttinger mechanism even for low
concentrations of charge carriers if we consider the
spin-polarized or two-band situation, as well as a multilayer
system. In this case, the role of the pairing spins up is played
by electrons of the first band (layer), while the role of the
screening spins down is played by electrons of the second band
(layer). Coupling between the electrons of the two bands occurs
via interband (interlayer) Coulomb interaction. As a result, the
following superconductivity mechanism becomes possible: electrons
of one species form a Cooper pair by polarizing electrons of
another species~\cite{Kagan91,KaganValkov11}. This mechanism of
superconductivity is also effective in quasi-two-dimensional
systems. Note that odd-momentum pairing and superconductivity in
vertical graphene heterostructures made up by graphene layers
separated by boron nitride spaces was considered recently by
Guinea and Uchoa~\cite{Uchoa07}.

In this work, we investigate the Kohn-Luttinger
Cooper instability in an idealized monolayer and
bilayer of doped graphene using the Born weak-coupling
approximation and taking into account the Coulomb
repulsion between electrons of the same and of
the nearest carbon atoms in a monolayer, as well as the
interlayer Coulomb repulsion in the case of the bilayer.

The necessity of including the long-range Coulomb interaction in
calculating the physical characteristics of graphene is dictated
by the results obtained in ~\cite{Wehling11}, where the partly
screened frequency-dependent Coulomb interaction was calculated ab
initio in constructing the effective multiparticle model of
graphene and graphite. It was found that the intra-atomic
repulsion potential in graphene is $U=9.3\,\textrm{eV}$ (an
analogous estimate is given in~\cite{Levin74}), which contradicts
the intuitively expected small value of $U$ and weak-coupling
limit $U<W$. The calculations performed in~\cite{Wehling11} have
demonstrated the fundamental importance of taking into account the
nonlocal Coulomb interaction in graphene: the Coulomb repulsion of
electrons at neighboring sites amounts to $V=5.5\,\textrm{eV}$. It
should be noted that the values of $U$ and $V$ estimated by other
researchers (see, e.g.,~\cite{Perfetto07}) are much smaller.

\section{IDEALIZED GRAPHENE MONOLAYER}
\label{monolayer}

In the hexagonal lattice of graphene, each unit cell
contains two carbon atoms; therefore, the entire lattice
can be divided into two sublattices $A$ and $B$. In the
Shubin-Vonsovsky (extended Hubbard) model, the
Hamiltonian for the graphene monolayer taking into
account electron hoppings between the nearest and
next-to-nearest atoms, as well as the Coulomb repulsion
between electrons of the same atom and of adjacent
atoms in the Wannier representation, has the form
\begin{eqnarray}\label{grapheneHamiltonian}
\hat{H}&=&\hat{H}_0+\hat{H}_{\textrm{int}},\\
\hat{H}_0&=&-\mu\Biggl(\sum_{f\sigma}\hat{n}^A_{f\sigma}+\sum_{g\sigma}
\hat{n}^B_{g\sigma}\Biggr)-\\
&-&t_1\sum_{
f\delta\sigma}(a^{\dag}_{f\sigma}b_{f+\delta,\sigma}+\textrm{h.c.})-\nonumber\\
&-&t_2\Biggl(\sum_{\langle\langle
fm\rangle\rangle}a^{\dag}_{f\sigma}a_{m\sigma}+\sum_{\langle\langle
gn\rangle\rangle}b^{\dag}_{g\sigma}b_{n\sigma}+
\textrm{h.c.}\Biggr),\nonumber\\\label{H0}
\hat{H}_{\textrm{int}}&=&U\Biggl(\sum_f
\hat{n}^{A}_{f\uparrow}\hat{n}^{A}_{f\downarrow}+\sum_g
\hat{n}^{B}_{g\uparrow}\hat{n}^{B}_{g\downarrow}\Biggr)+\nonumber\\
&+&V\sum_{ f\delta\sigma}
\hat{n}^{A}_{f\sigma}\hat{n}^{B}_{f+\delta,\sigma}.\label{Hint}
\end{eqnarray}
Here, operators $a^{\dag}_{f\sigma}(a_{f\sigma})$ create (annihilate) an
electron with spin projection $\sigma=\pm1/2$ at site $f$ of sublattice
$A$; $\displaystyle\hat{n}^{A}_{f\sigma}=a^{\dag}_{f\sigma}a_{f\sigma}$
denotes the operators of the number of
fermions at the $f$ site of sublattice $A$ (analogous notation
is used for sublattice $B$). Vector $\delta$ connects the
nearest atoms of the hexagonal lattice. In the Hamiltonian,
the symbol $\langle\langle~\rangle\rangle$ indicates that summation is
carried out only over next-to-nearest neighbors.

Passing to the momentum space and performing
the Bogoliubov transformation,
\begin{eqnarray}\label{uv}
\alpha _{i\textbf{k}\sigma}= w_{i1}(\textbf{k}){a_{
\textbf{k}\sigma }} + w_{i2}(\textbf{k}){b_{\textbf{k}\sigma
}},\qquad i=1,2,
\end{eqnarray}
we diagonalize Hamiltonian $\hat{H}_0$, which acquires the
form
\begin{eqnarray}
\hat H_0 =\sum\limits_{i=1}^2 \sum\limits_{ \textbf{k}\sigma }
E_{i\textbf{k}}
{\alpha_{i\textbf{k}\sigma}^{\dag}\alpha_{i\textbf{k}\sigma}}.
\end{eqnarray}
The two-band energy spectrum is described by the
expressions~\cite{Wallace47}
\begin{eqnarray}\label{spectra}
E_{1\textbf{k}}=t_1|u_{\textbf{k}}|-t_2f_{\textbf{k}},\qquad
E_{2\textbf{k}}=-t_1|u_{\textbf{k}}|-t_2f_{\textbf{k}},
\end{eqnarray}
where the following notation has been introduced:
\begin{eqnarray}\label{f_k}
&&f_{\textbf{k}}=2\cos(\sqrt{3}k_y)+
4\cos\biggl(\frac{\sqrt{3}}{2}k_y\biggr)\cos\biggl(\frac{3}{2}k_x\biggr),\\
&&u_{\textbf{k}}=\displaystyle\sum_{\delta}e^{i
\textbf{k}\delta}=e^{-ik_x}+
2e^{\frac{i}{2}k_x}\cos\biggl(\frac{\sqrt{3}}{2}k_y\biggr),\label{u_k}\\
&&|u_{\textbf{k}}|=\sqrt{3+f_{\textbf{k}}}.
\end{eqnarray}
The coefficients of the Bogoliubov transformation
have the form
\begin{eqnarray}\label{wz}
&&{w_{1,1}}(\textbf{k}) = w_{22}^*(\textbf{k}) =
\frac{1}{\sqrt{2}}r_\textbf{k}^*,
\qquad r_\textbf{k}=\frac{u_\textbf{k}}{|u_\textbf{k}|},\\
&&{w_{12}}(\textbf{k}) = -w_{21}(\textbf{k})
=-\frac{1}{\sqrt{2}}.\nonumber
\end{eqnarray}

In the Bogoliubov representation, interaction
operator (\ref{Hint}) is defined by the following expression
including operators $\alpha_{1\textbf{k}\sigma}$ and $\alpha_{2\textbf{k}\sigma}$:
$\alpha_{1\textbf{k}\sigma}$ è $\alpha_{2\textbf{k}\sigma}$
\begin{eqnarray}\label{Hint_ab}
\hat H_{int} &=& \frac{1}{N}\sum\limits_{ijlm\atop
\textbf{k}\textbf{p}\textbf{q}\textbf{s}\sigma}
\Gamma_{ij;lm}^{||}(\textbf{k},\textbf{p}|\textbf{q},\textbf{s})
\alpha_{i\textbf{k}\sigma}^\dag \alpha_{j\textbf{p}\sigma}^\dag
\alpha_{l\textbf{q}\sigma}\alpha_{m\textbf{s}\sigma}
\times\nonumber\\
&\times&\Delta (\textbf{k}+\textbf{p}-\textbf{q}-\textbf{s}) +\\
&+& \frac{1}{N}\sum\limits_{ijlm\atop
\textbf{k}\textbf{p}\textbf{q}\textbf{s}}\Gamma_{ij;lm}^{\bot}(\textbf{k},\textbf{p}|\textbf{q},\textbf{s})
\alpha_{i\textbf{k}\uparrow}^\dag
\alpha_{j\textbf{p}\downarrow}^\dag \alpha_{l\textbf{q}\downarrow}
\alpha_{m\textbf{s}\uparrow}\times\nonumber\\
&\times&\Delta
(\textbf{k}+\textbf{p}-\textbf{q}-\textbf{s}),\nonumber
\end{eqnarray}
where $\Delta$ is the Kronecker symbol, while $\Gamma_{ij;lm}^{||}(\textbf{k},\textbf{p}|\textbf{q},\textbf{s})$
and $\Gamma_{ij;lm}^{\bot}(\textbf{k},\textbf{p}|\textbf{q},\textbf{s})$ are the initial amplitudes. The
quantity
\begin{eqnarray}
\Gamma_{ij;lm}^{||}(\textbf{k},\textbf{p}|\textbf{q},\textbf{s})
&=& \frac
12\Bigl(V_{ij;lm}(\textbf{k},\textbf{p}|\textbf{q},\textbf{s})+\nonumber\\
&+&V_{ji;ml}(\textbf{p},\textbf{k}|\textbf{s},\textbf{q})\Bigr),\\
V_{ij;lm}(\textbf{k},\textbf{p}|\textbf{q},\textbf{s})&=&V
{u_{\textbf{q}-\textbf{p}}} w_{i1}(\textbf{k}) w_{j2}(\textbf{p})
w^*_{l2}(\textbf{q}) w^*_{m1}(\textbf{s})\nonumber\\
\end{eqnarray}
describes the intensity of the interaction of fermions
with parallel spin projections, while the quantity
\begin{eqnarray}
&&\Gamma_{ij;lm}^{\bot}(\textbf{k},\textbf{p}|\textbf{q},\textbf{s})
=U_{ij;lm}(\textbf{k},\textbf{p}|\textbf{q},\textbf{s})
+\nonumber\\
&&\qquad\qquad+V_{ij;lm}(\textbf{k},\textbf{p}|\textbf{q},\textbf{s})+
V_{ji;ml}(\textbf{p},\textbf{k}|\textbf{s},\textbf{q}),\\
&&U_{ij;lm}(\textbf{k},\textbf{p}|\textbf{q},\textbf{s}) =
U\Bigl( w_{i1}(\textbf{k}) w_{j1}(\textbf{p}) w^*_{l1}(\textbf{q}) w^*_{m1}(\textbf{s}) +\nonumber\\
&&\qquad\qquad+w_{i2}(\textbf{k}) w_{j2}(\textbf{p})
w^*_{l2}(\textbf{q}) w^*_{m2}(\textbf{s}) \Bigr)
\end{eqnarray}
corresponds to the interaction of Fermi quasiparticles
with antiparallel spin projections. Indices ${i,j,l,m}$ can
acquire values of 1 or 2.

\begin{figure*}
\begin{center}
\includegraphics[width=0.98\textwidth]{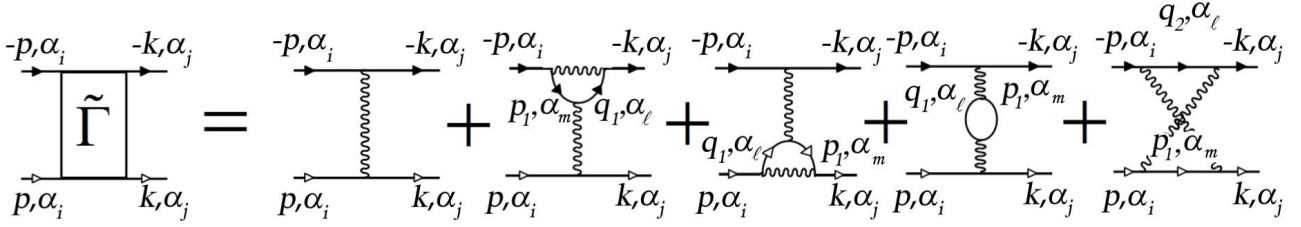}
\caption{Fig. 1. First- and second-order diagrams for the
effective interaction of electrons in graphene monolayer and
bilayer. Solid lines with light (dark) arrows correspond to the
Green's functions for electrons with spin projections
$+{\textstyle{1 \over2}}$~($-{\textstyle{1 \over 2}}$) and
energies corresponding to graphene energy bands $\alpha_i$,
$\alpha_j$, $\alpha_l$ and $\alpha_m$. In diagrams for the
monolayer (Section 2), subscripts $i=j=1$; subscripts $l$ and $m$
can acquire values of 1 or 2. In the case of the bilayer (Section
3), subscripts $i$ and $j$ acquire values of 1 or 2, while
subscripts $l$ and $m$ acquire values of 1, 2, 3, or 4. Momenta
$q_1$ and $q_2$ are defined by relations (\ref{q1q2}).}
\label{diagrams_alpha}
\end{center}
\end{figure*}

Using the Born weak coupling approximation
(with the hierarchy $W>U>V$ of the model parameters,
where $W$ is the bandwidth for the upper and lower
branches of the energy spectrum (\ref{spectra}) and (\ref{f_k}) of
graphene for the case of $t_2 =0$), we can consider the
scattering amplitude in the Cooper channel, confining
our analysis to only second-order diagrams in the
effective interaction of two electrons with opposite
values of the momentum and spin and using quantity
$\widetilde{\Gamma}(\textbf{p}, \textbf{k})$ for it.
Graphically, this quantity is determined
by the sum of the diagrams shown in Fig.~\ref{diagrams_alpha}. Solid lines
with light (dark) arrows correspond to Green's functions
for electrons with opposite values of spin projections $+\frac12 ~(-\frac12)$.
It is well known that the possibility of
Cooper pairing is determined by the characteristics of
the energy spectrum near the Fermi level and by the
effective interaction of electrons located near the Fermi
surface~\cite{Gor'kov61}. Assuming that the chemical potential in
doped graphene falls into the upper energy band $E_{1\textbf{k}}$ and
analyzing the conditions for the occurrence of Kohn-
Luttinger superconductivity, we can consider the situation
in which the initial and final momenta also belong
to the upper subband. This is reflected in Fig.~\ref{diagrams_alpha} via indices
$\alpha_1$ (upper band) and $\alpha_2$ (lower band).

The first diagram in Fig.~\ref{diagrams_alpha} corresponds to the initial
interaction of two electrons in the Cooper channel.
The next (Kohn-Luttinger) diagrams in Fig.~\ref{diagrams_alpha}
describe second-order scattering processes, $\delta\widetilde{\Gamma}(\textbf{p}, \textbf{k})$,
and take into account the polarization effects of the
filled Fermi sphere. Two solid lines without arrows in
these diagrams indicate summation over both spin
projections. Wavy lines correspond to the initial interaction.
Scattering of electrons with identical spin projections
corresponds only to the intersite contribution.
If electrons with different spin projections interact, the
scattering amplitude is determined by the sum of the
Hubbard and intersite repulsions. Thus, in the presence
of the short-range Coulomb interaction alone,
the correction $\delta\widetilde{\Gamma}(\textbf{p}, \textbf{k})$
 to the effective interaction is
determined by the last exchange diagram only. If the
Coulomb interaction of electrons at neighboring lattice
sites of graphene is taken into account, all diagrams in
Fig.~\ref{diagrams_alpha} contribute to the renormalized amplitude.

After the introduction of the analytical expressions
for the diagrams, we obtain the following analytic
expression for the effective interaction in Fig.~\ref{diagrams_alpha}:
\begin{eqnarray}\label{Gamma_wave}
&&\widetilde{\Gamma}(\textbf{p},\textbf{k})= \frac U2+ \frac
V2\textrm{Re}(u_{\textbf{p}-\textbf{k}} r^*_{\textbf{p}}
r_{\textbf{k}})+\delta\widetilde{\Gamma}(\textbf{p},\textbf{k}),\\
\text{where}\nonumber\\
&&\delta \tilde \Gamma (\textbf{p},\textbf{k})=
\frac{1}{N}\sum\limits_{i,j, \textbf{p}_1}
\Gamma^{\bot}_{1i;1j}(\textbf{p}, \textbf{q}_2| -\textbf{k},
\textbf{p}_1)
\times \\
&&\times\Gamma^{\bot}_{j1;i1}(\textbf{p}_1,
-\textbf{p}|\textbf{q}_2,\textbf{k})
\chi_{i,j}(\textbf{q}_2, \textbf{p}_1)+\nonumber\\
&&+\frac{2}{N}\sum\limits_{i,j, \textbf{p}_1}  \Bigl\{
\Gamma^{\bot}_{1j;i1}( \textbf{p}, \textbf{p}_1| \textbf{q}_1,
\textbf{k})\times\nonumber\\
&&\times\left[\Gamma^{||}_{i1;j1}( \textbf{q}_1, -\textbf{p}|
\textbf{p}_1, -\textbf{k}) -
\Gamma^{||}_{i1;1j}( \textbf{q}_1, -\textbf{p}| -\textbf{k}, \textbf{p}_1) \right] \Bigr.+\nonumber\\
&&+\Bigl.\Gamma^{\bot}_{i1;1j}(\textbf{q}_1, -\textbf{p}| -\textbf{k}, \textbf{p}_1)\times\nonumber\\
&&\times\left[\Gamma^{||}_{1j;1i}( \textbf{p}, \textbf{p}_1|
\textbf{k}, \textbf{q}_1) - \Gamma^{||}_{1j;i1}( \textbf{p},
\textbf{p}_1| \textbf{q}_1, \textbf{k})\right]
\Bigr\}\chi_{i,j}(\textbf{q}_1,\textbf{p}_1).\nonumber
\end{eqnarray}
Here, we have introduced the following notation for
generalized susceptibilities:
\begin{equation}
\chi_{i,j}(\textbf{k},\textbf{p}) = \frac{f(E_{i\textbf{k}}) -
f(E_{j\textbf{p}})} {E_{j\textbf{p}} - E_{i\textbf{k}}},
\end{equation}
where
\[f(x)=(\exp(\frac{x-\mu}{T})+1)^{-1}\]
--- is the Fermi-Dirac distribution function, and energies $E_{i\textbf{k}}$
are defined by expressions~(\ref{spectra}). For the sake of
compactness, we have introduced the notation for the
combinations of momenta:
\begin{equation}\label{q1q2}
\textbf{q}_1 =  \textbf{p}_1 + \textbf{p} - \textbf{k},\qquad
\textbf{q}_2 = \textbf{p}_1-\textbf{p}-\textbf{k}.
\end{equation}

Knowing the renormalized expression for the
effective interaction, we can pass to analysis of the
conditions for the emergence of superconductivity in
the system under investigation. It is well known~\cite{Gor'kov61}
that the emergence of Cooper instability can be established
from analysis of the homogeneous part of the
Bethe-Saltpeter equation. In this case, the dependence
of the scattering amplitude $\Gamma(\textbf{p},\textbf{k})$  on momentum $\textbf{k}$
can be factorized, which gives the integral equation
for the superconducting order parameter $\Delta(\textbf{p})$.
After integrating over isoenergetic contours, we can
reduce the problem of the Cooper instability to the
eigenvalue problem~\cite{Scalapino86,Baranov92,Hlubina99,Raghu10,Alexandrov11,Kagan13}
\begin{equation}
\label{IntegralEqPhi}
\frac{3\sqrt{3}}{8\pi^2}\oint\limits_{\varepsilon_{\textbf{q}}=\mu}
\frac{d\hat{\textbf{q}}} {v_F(\hat{\textbf{q}})}
\widetilde{\Gamma}(\hat{\textbf{\textbf{p}}},\hat{\textbf{q}})
\Delta(\hat{\textbf{q}})=\lambda\Delta(\hat{\textbf{p}}),
\end{equation}
where superconducting order parameter $\Delta(\hat{\textbf{q}})$ plays the
role of the eigenvector, and eigenvalues $\lambda$ satisfy the
relation $\lambda^{-1}\simeq \ln(T_c/W)$. In this case, momenta $\hat{\textbf{p}}$ and
$\hat{\textbf{q}}$ lie on the Fermi surface and $v_F(\hat{\textbf{q}})$ is the Fermi velocity.

To solve Eq. (\ref{IntegralEqPhi}), we write its kernel as the superposition
of eigenfunctions each of which belongs to
one of irreducible representations of symmetry group
$C_{6v}$ of the hexagonal lattice. It is well known that this
symmetry group has six irreducible representations
~\cite{Landau89}: four 1D and two 2D representations. For each
representation, Eq. (\ref{IntegralEqPhi}) has a solution with its own
effective coupling constant ë. We will henceforth use
the following notation for the classification of the order
parameter symmetries: representation $A_1$ corresponds
to the $s-$wave symmetry type; $B_1$ and $B_2$ correspond to
the $f-$wave symmetry; $E_1$, to the $p + ip-$wave symmetry
type; and $E_2$, to the $d + id-$wave symmetry type.

For each irreducible representation $\nu$, we will seek
the solution to Eq. (\ref{IntegralEqPhi}) in the form
\begin{equation}\label{solution}
\Delta^{(\nu)}(\phi)=\sum\limits_{m}\Delta_{m}^{(\nu)}g_{m}^{(\nu)}(\phi),
\end{equation}
where $m$ is the number of the eigenfunction belonging
to representation $\nu$ and $\phi$ is the angle determining the
direction of momentum $\hat{\textbf{p}}$ relative to the $p_x$ axis. The
explicit form of the orthonormal functions $g_{m}^{(\nu)}(\phi)$ is
defined by the expressions
\begin{eqnarray}\label{harmon}
A_1&&\rightarrow~g_{m}^{(s)}(\phi)=\frac{1}{\sqrt{(1+\delta_{m0})\pi}}\,
\textrm{cos}\,6m\phi,~~m\in[\,0,\infty),\label{invariants_s}\nonumber\\
A_2&&\rightarrow~g_{m}^{(A_2)}(\phi)=\frac{1}{\sqrt{\pi}}\,\textrm{sin}\,
(6m+6)\phi,\label{invariants_s1}\nonumber\\
B_1&&\rightarrow~g_{m}^{(f_1)}(\phi)=\frac{1}{\sqrt{\pi}}\,
\textrm{sin}\,(6m+3)\phi,\label{invariants_dxy}\\
B_2&&\rightarrow~g_{m}^{(f_2)}(\phi)=\frac{1}{\sqrt{\pi}}\,
\textrm{cos}\,(6m+3)\phi,\label{invariants_dx2y2}\nonumber\\
E_1&&\rightarrow~g_{m}^{(p+ip)}(\phi)=\frac{1}{\sqrt{\pi}}\,(A\,\textrm{sin}\,
(2m+1)\phi+\nonumber\\
&&\qquad+B\,\textrm{cos}\,(2m+1)\phi),\label{invariants_p}\nonumber\\
E_2&&~\rightarrow~g_{m}^{(d+id)}(\phi)=\frac{1}{\sqrt{\pi}}\,(A\,\textrm{sin}\,
(2m+2)\phi+\nonumber\\
&&\qquad+B\,\textrm{cos}\,(2m+2)\phi)\label{invariants_p}\nonumber.
\end{eqnarray}
Here, subscripts $m$ for the 2D representations $E_1$ and $E_2$
run through the values for which coefficients $(2m + 1)$
and $(2m + 2)$, respectively, are not multiples of three.

The basis functions satisfy the orthonormality conditions
\begin{equation}\label{norma}
\int\limits_0^{2\pi}d\phi\,g_{m}^{(\nu)}(\phi)g_{
n}^{(\beta)}(\phi)=\delta_{\nu\beta}\delta_{mn}.
\end{equation}

Substituting expression (\ref{solution}) into Eq. (\ref{IntegralEqPhi}),
integrating with respect to angles, and using condition (\ref{norma}), we
obtain
\begin{equation}\label{EqDelta}
\sum_n\Lambda^{(\nu)}_{mn}\Delta_{n}^{(\nu)}=\lambda_{\nu}\Delta_{m}^{(\nu)},
\end{equation}
where
\begin{eqnarray}
\label{matrix}
\Lambda^{(\alpha)}_{mn}&=&\frac{3\sqrt{3}}{8\pi^2}\oint\limits_0^{2\pi}d\phi_{\textbf{p}}
\oint\limits_0^{2\pi}d\phi_{\textbf{q}}\frac{d\hat{\textbf{q}}}
{d\phi_{\textbf{q}}v_F(\hat{\textbf{q}})}
\widetilde{\Gamma}({\hat{\textbf{p}}}\,|{\hat{\textbf{q}}})\times\nonumber\\
&\times&g_{m}^{(\nu)}(\phi_{\textbf{p}}) g_{
n}^{(\nu)}(\phi_{\textbf{q}}).
\end{eqnarray}
Since $T_c\sim W\exp\bigl(1/\lambda \bigr)$, each negative eigenvalue $\lambda_{\nu}$
corresponds to the superconducting phase with the
order parameter symmetry $\nu$. Generally speaking, the
expansion of the order parameter $\Delta^{(\nu)}(\phi)$ in eigenfunctions
includes a large number of harmonics; however,
the main contribution is determined by only some of
these harmonics. The highest value of the superconducting
transition temperature corresponds to the
modulus of the largest value of $\lambda_{\nu}$.

\begin{figure}[t]
\begin{center}
\includegraphics[width=0.47\textwidth]{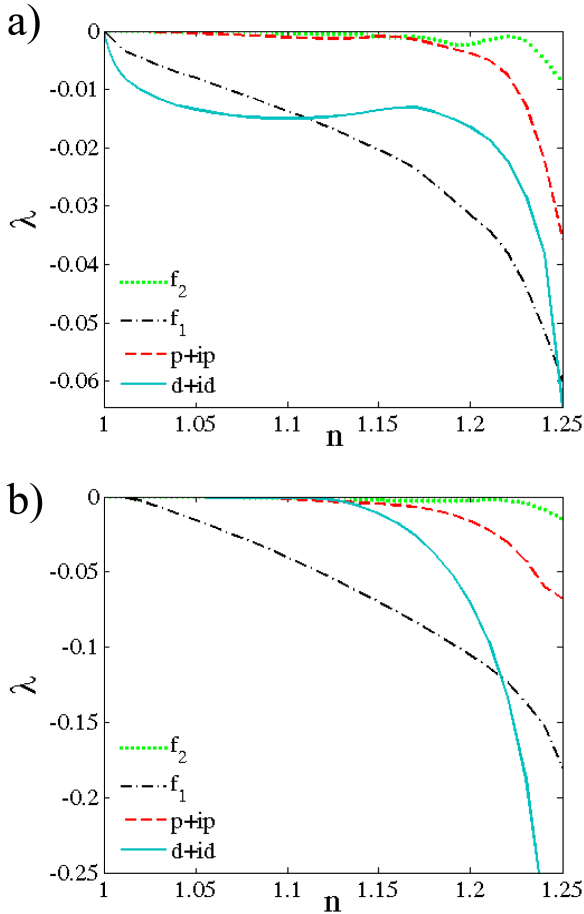}
\caption{Fig. 2. Dependences of $\lambda$ on carrier concentration
$n$ in the graphene monolayer: (a) $t_2 = 0, U = 2|t_1|$, and $V =
0$; (b) $t_2 = 0, U = 2|t_1|$, and $V =
0.5|t_1|$.}\label{lambdas_mlg}
\end{center}
\end{figure}

Figure~\ref{lambdas_mlg}a shows the calculated dependencies of the
effective coupling constant $\lambda$  on carrier concentration
$n$ for various symmetry types of the superconducting
order parameter for the set of parameters $t_2=0,\,U=2|t_1|$,
and $V = 0$. It can be seen that for low electron
densities $1 < n < 1.12$, in the vicinity of the van Hove
singularity, the competition occurs between the superconducting
phases with the $f-$wave symmetry type,
whose contribution is determined by the harmonics
$g_{m}^{(f_1)}(\phi)=\displaystyle\frac{1}{\sqrt{\pi}}\,
\textrm{sin}\,(6m+3)\phi$, and the $d + id-$wave symmetry type corresponding to 2D
representation $E_2$. In the electron concentration range
$1 < n < 1.12$, the $d + id-$wave pairing prevails, while for
$1.12<n<n_{vH}$, superconductivity with the $f-$wave symmetry
type of the order parameter is stabilized.

\begin{figure}[t]
\begin{center}
\includegraphics[width=0.47\textwidth]{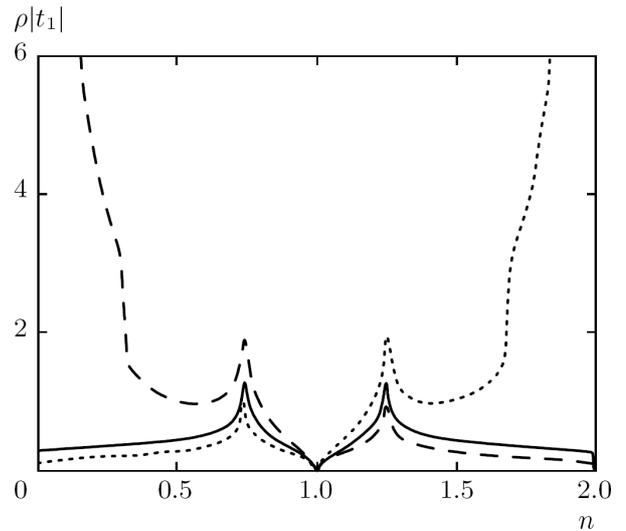}
\caption{Fig. 3. Modification of the electron density of states
for the graphene monolayer upon switching of the hoppings to the
next-to-nearest atoms for $t_2 =0$ solid curve), $t_2 = -0.2|t_1|$
(dashed curve), and $t_2 = 0.2|t_1|$ (dotted
curve).}\label{DOS_SLG}
\end{center}
\end{figure}

It should be noted that to avoid the summation of
parquet diagrams~\cite{Dzyaloshinskii88,Zheleznyak97}, we do not analyze here the
electron concentration ranges which are too close
to the van Hove singularity (Fig.~\ref{DOS_SLG}).

The account of the intersite Coulomb interaction considerably
affects the competition between superconducting phases. This can
be seen from Fig.~\ref{lambdas_mlg}b which shows the $\lambda(n)$
dependences for parameters $t_2 = 0, U = 2|t_1|$, and $V =
0.5|t_1|$. Comparison with Fig.~\ref{lambdas_mlg}a shows that the
switching of the intersite Coulomb interaction suppresses Cooper
pairing in the $d + id$-wave channel for low electron densities;
however, it leads to superconductivity with this type of symmetry
in the vicinity of the van Hove singularity. As a result, the
$f$-wave pairing takes place in the electron concentration range
$1 < n < 1.22$. It should be noted that this result is in
qualitative agreement with the results obtained
in~\cite{Kiesel12}.

The switching of electron hoppings $t_2$ to the next-to-nearest
carbon atoms in the graphene monolayer does not qualitatively
affect the competition between the superconducting phases of
different symmetry types, which is illustrated in
Fig.~\ref{lambdas_mlg}b~\cite{Kagan14}. Such a behavior of the
system can be explained by the fact that an account of hoppings
$t_2> 0$ or $t_2 < 0$ does not noticeably modify the density of
electron states of the monolayer in the range of carrier
concentrations between the Dirac point and both van Hove
singularities (Fig.~\ref{DOS_SLG}). However, the inclusion of
hoppings $t_2$ leads to an increase in the absolute values of the
effective interaction and, hence, to higher superconducting
transition temperatures in the idealized graphene
monolayer~\cite{Kagan14}.

\section{IDEALIZED GRAPHENE BILAYER}
\label{bilayer}
\begin{figure}[t]
\begin{center}
\includegraphics[width=0.47\textwidth]{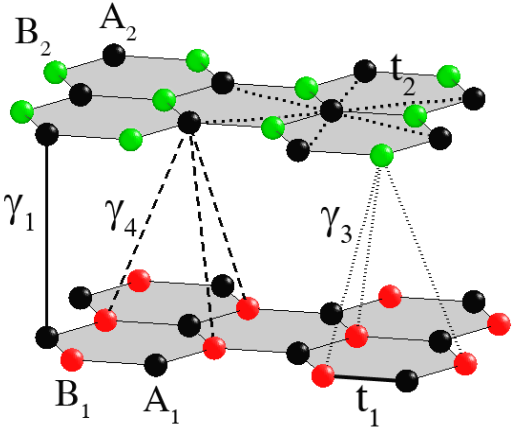}
\caption{Fig. 4. Crystal structure of the graphene bilayer. Carbon
atoms $A_1$ and $B_1$ in the lower layer are shown by red and
black balls; atoms $A_2$ and $B_2$ in the upper layer are shown by
black and green balls. Intralayer electron hoppings are marked by
$t_1$ and $t_2$; $\gamma_1,\,\gamma_3$, and $\gamma_4$ show the
interplanar hoppings.}\label{bilayer_structure}
\end{center}
\end{figure}

Let us consider an idealized graphene bilayer, assuming that two
monolayers are arranged in accordance with the $AB$ type (i.e.,
one monolayer is turned through 60$^o$ relative to the other
monolayer)~\cite{McCann06,McCann13}. We choose the arrangement of
the sublattices in the layers in such a way that the sites from
different layers located one above another belong to sublattice
$A$, while the remaining sites belong to sublattice $B$
(Fig.~\ref{bilayer_structure}). In this case, the Hamiltonian of
the graphene bilayer in the Wannier representation has the form
\begin{eqnarray}\label{HamiltonianBilayer}
\hat{H}&=&\hat{H}_0+\hat{H}_{\textrm{int}},\\
\hat{H}_0&=&(\varepsilon-\mu)\Biggl(\sum_{if\sigma}\hat{n}^{A}_{if\sigma}+
\sum_{ig\sigma}\hat{n}^{B}_{ig\sigma}\Biggr)\nonumber\\
&-&t_1\sum_{f\delta\sigma}(a^{\dag}_{1f\sigma}b_{1,f+\delta,\sigma}+
a^{\dag}_{2f\sigma}b_{2,f-\delta,\sigma}+\textrm{h.c.})\nonumber\\
&-&t_2\sum_{i\sigma}\Biggl(\sum_{\langle\langle
fm\rangle\rangle}a^{\dag}_{if\sigma}a_{im\sigma}+\sum_{\langle\langle
gn\rangle\rangle}b^{\dag}_{ig\sigma}b_{in\sigma}+
\textrm{h.c.}\Biggr)\nonumber\\
&-&\gamma_1\sum_{f\sigma}(a^{\dag}_{1f\sigma}a_{2f\sigma}+\textrm{h.c.})-\gamma_3\sum_{g\delta\sigma}(b^{\dag}_{1g\sigma}b_{2,g+\delta,\sigma}+\textrm{h.c.})\nonumber\\
&-&\gamma_4\sum_{f\delta\sigma}(a^{\dag}_{1f\sigma}b_{2,f-\delta,\sigma}+
a^{\dag}_{2f\sigma}b_{1,f+\delta,\sigma}+\textrm{h.c.}),\label{H0Bilayer}\\
\hat{H}_{\textrm{int}}&=&U\biggl(\sum_{if}
\hat{n}^{A}_{if\uparrow}\hat{n}^{A}_{if\downarrow}+ \sum_{ig}
\hat{n}^{B}_{ig\uparrow}\hat{n}^{B}_{ig\downarrow}\biggr)\nonumber\\
&+&V\sum_{f\delta\sigma}
\Bigl(\hat{n}^{A}_{1f\sigma}\hat{n}^{B}_{1,f+\delta,\sigma}+
\hat{n}^{A}_{2f\sigma}\hat{n}^{B}_{2,f-\delta,\sigma}\Bigr)\nonumber\\
&+&G_1\sum_{f\sigma} \hat{n}^{A}_{1f\sigma}\hat{n}^{A}_{2f\sigma}
+ G_3\sum_{g\delta\sigma}
\hat{n}^{B}_{1g\sigma}\hat{n}^{B}_{2,g+\delta,\sigma}\nonumber\\
&+&G_4\sum_{f\delta\sigma}
\Bigl(\hat{n}^{A}_{1f\sigma}\hat{n}^{B}_{2,f-\delta,\sigma}+
\hat{n}^{A}_{2f\sigma}\hat{n}^{B}_{1,f+\delta,\sigma}\Bigr).\label{HintBilayer}
\end{eqnarray}
Here, we have used notation analogous to that for Hamiltonian
(\ref{grapheneHamiltonian}) for a monolayer in Section 2. Index
$i=1,2$ in Hamiltonian (\ref{HamiltonianBilayer}) denotes the
number of the monolayer. We assume that one-site energies are
identical
($\varepsilon_{A1}=\varepsilon_{A2}=\varepsilon_{B1}=\varepsilon_{B2}=\varepsilon$).
Interlayer electron hopping parameters are denoted by
$\gamma_1,\,\gamma_3,\,\gamma_4$ (see
Fig.~\ref{bilayer_structure}). The last three terms in Hamiltonian
(\ref{HamiltonianBilayer}) take into account the interlayer
Coulomb interaction of electrons in atoms $A_1$ and $A_2$, $B_1$
and $B_2$, and $A_1$ and $B_2$; the intensities of these
interactions are denoted by $G_1$, $G_3$, and $G_4$, respectively.

Passing to the momentum space, it is convenient to write
Hamiltonian $\hat{H}_0$ in matrix form:
\begin{eqnarray}\label{HamiltonianBilayerMatrix}
&&\hat{H}_0=\\
&&-\sum_{\textbf{k}\sigma}
\left(
\begin{array}{c}
  a^{\dag}_{1\textbf{k}\sigma} \\
  a^{\dag}_{2\textbf{k}\sigma} \\
  b^{\dag}_{1\textbf{k}\sigma} \\
  b^{\dag}_{2\textbf{k}\sigma} \\
\end{array}
\right)^T
\left(
\begin{array}{cccc}
  \varepsilon_{\textbf{k}} & \gamma_1 & t_1u^*_{\textbf{k}} & \gamma_4u_{\textbf{k}} \\
  \gamma_1 & \varepsilon_{\textbf{k}} & \gamma_4u^*_{\textbf{k}} & t_1u_{\textbf{k}} \\
  t_1u_{\textbf{k}} & \gamma_4u_{\textbf{k}} & \varepsilon_{\textbf{k}} & \gamma_3u^*_{\textbf{k}} \\
  \gamma_4u^*_{\textbf{k}} & t_1u^*_{\textbf{k}} & \gamma_3u_{\textbf{k}} & \varepsilon_{\textbf{k}} \\
\end{array}
\right)
\left(
\begin{array}{c}
  a_{1\textbf{k}\sigma} \\
  a_{2\textbf{k}\sigma} \\
  b_{1\textbf{k}\sigma} \\
  b_{2\textbf{k}\sigma} \\
\end{array}
\right),\nonumber
\end{eqnarray}
where $\varepsilon_{\textbf{k}}=t_2f_{\textbf{k}}-\varepsilon$,
and quantity $f_{\textbf{k}}$ is defined by expression
(\ref{f_k}).

Hamiltonian $\hat{H}_0$ can be diagonalized using the Bogoliubov
transformation
\begin{eqnarray}\label{uv2}
\alpha_{i\textbf{k}\sigma}&=& w_{i1}(\textbf{k}){a_{1
\textbf{k}\sigma }} + w_{i2}(\textbf{k}){a_{2\textbf{k}\sigma
}}+\\
&+&w_{i3}(\textbf{k}){b_{1\textbf{k}\sigma }} +
w_{i4}(\textbf{k}){b_{2\textbf{k}\sigma }}.\quad
i=1,2,3,4,\nonumber
\end{eqnarray}
It acquires the form
\begin{eqnarray}
\hat H_0 =\sum\limits_{i=1}^4 \sum\limits_{ \textbf{k}\sigma }
E_{i\textbf{k}}
{\alpha_{i\textbf{k}\sigma}^{\dag}\alpha_{i\textbf{k}\sigma}}.
\end{eqnarray}
According to the results of \cite{Dresselhaus02,Brandt88}, the
interlayer hoppings $\gamma_4$ are relatively weak, so it allows
us to assume that $\gamma_4=0$ for convenience of diagonalization
of the Hamiltonian. In this case, the four-band energy spectrum of
the graphene bilayer is described by the expressions
\begin{eqnarray}
&&E_{i\textbf{k}}=\varepsilon\pm\sqrt{A_{\textbf{k}}\pm\sqrt{B_{\textbf{k}}}}-t_2f_{\textbf{k}},\\
&&A_{\textbf{k}}=\frac14\Bigl(2a^2+4|b_{\textbf{k}}|^2+2|d_{\textbf{k}}|^2\Bigr),\nonumber\\
&&B_{\textbf{k}}=\frac14\Bigl(|d_{\textbf{k}}|^2
(|d_{\textbf{k}}|^2-2a^2+4|b_{\textbf{k}}|^2)+a^4+4a^2|b_{\textbf{k}}|^2+\nonumber\\
&&\qquad+4ab^{*2}_{\textbf{k}}d^*_{\textbf{k}}+4ab_{\textbf{k}}^2d_{\textbf{k}}\Bigr),\nonumber\\
&&a=\gamma_1,\quad b_{\textbf{k}}=t_1u_{\textbf{k}},\quad
d_{\textbf{k}}=\gamma_3u_{\textbf{k}}\nonumber,
\end{eqnarray}
where quantity $u_{\textbf{k}}$ is defined by expression
(\ref{u_k}).
\begin{figure}[t]
\begin{center}
\includegraphics[width=0.47\textwidth]{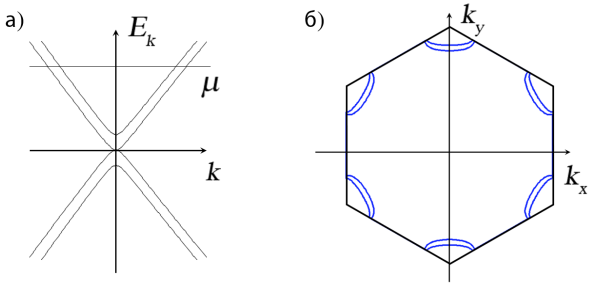}
\caption{Fig. 5. (a) Energy structure of the graphene bilayer near
Dirac points and (b) formation of the multisheet Fermi contour at
weak doping.}\label{two_contours}
\end{center}
\end{figure}

Analysis of the conditions for the occurrence of Kohn-Luttinger
superconductivity in the graphene bilayer is carried out in
accordance with the general scheme described in Section 2. We
consider the situation in which, as a result of doping, the
chemical potential of the bilayer is in the two upper energy bands
$E_{1\textbf{k}}$ and $E_{2\textbf{k}}$ as shown in
Fig.~\ref{two_contours}a. In this case, the initial and final
momenta of electrons in the Cooper channel also belong to the
upper two bands; for this reason, indices $i$ and $j$ in the
Kohn-Luttinger diagrams for a bilayer (see
Fig.~\ref{diagrams_alpha}) acquire the values 1 or 2. Writing the
analytical expressions for the diagrams, we obtain the analytic
expression for the effective interaction of electrons in the
Cooper channel of the graphene bilayer in
Fig.~\ref{diagrams_alpha}, which can subsequently be used for
analyzing the homogeneous part of the Bethe-Saltpeter equation.
When solving eigenvalue problem (\ref{IntegralEqPhi}), integration
is carried out with the allowance for the multisheet nature of
isoenergetic contours (Fig.~\ref{two_contours}b).

\begin{figure}[t]
\begin{center}
\includegraphics[width=0.47\textwidth]{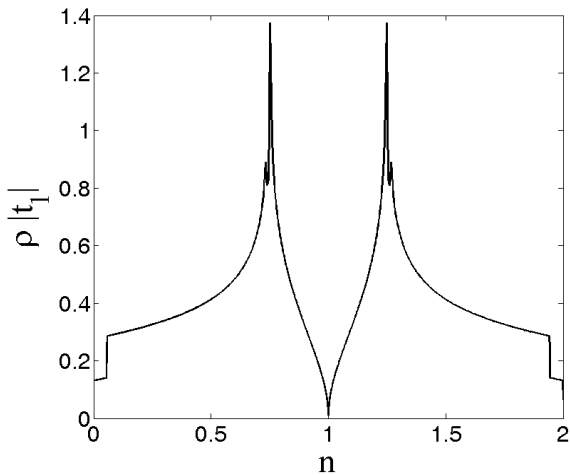}
\caption{Fig. 6. Dependence of the electron density of states for
the graphene bilayer per unit cell of one layer on the electron
concentration for the set of parameters
$t_2=0,\,\gamma_1=0.12|t_1|,\,\gamma_3=0.1|t_1|$.}\label{DOS_bilayer}
\end{center}
\end{figure}

\begin{figure}[t]
\begin{center}
\includegraphics[width=0.47\textwidth]{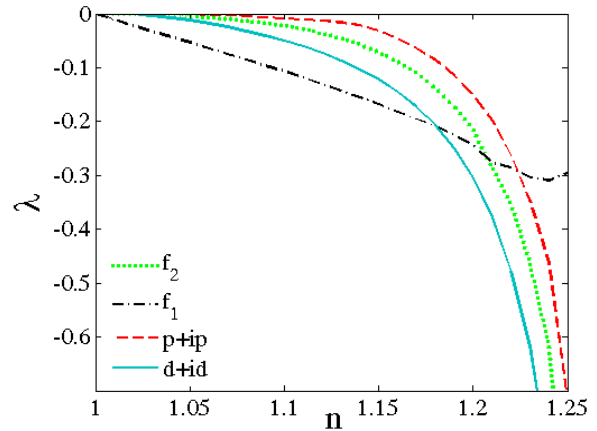}
\caption{Fig. 7. Dependence of $\lambda$ on carrier concentration
$n$ in the graphene bilayer for the set of parameters
$t_2=0,\,U=2,\,\gamma_1=0.12,\,\gamma_3=0.1,\,
V=0.5,\,G_1=G_3=G_4=0.5$ (all parameters are given in the units of
$|t_1|$).}\label{lambdas_blg}
\end{center}
\end{figure}

Let us now consider the dependences of effective coupling constant
$\lambda$ on carrier concentration $n$ for various types of
symmetry of the superconducting order parameter in the graphene
bilayer. It should be noted that in numerical calculations for the
graphene bilayer for $\gamma_1=\gamma_3=\gamma_4=0$ and
$G_1=G_3=G_4=0$, we get a limiting transition to the results
obtained in Section 2 for a graphene monolayer.
Figure~\ref{lambdas_blg} shows the $\lambda(n)$ dependences
determined for the bilayer with the set of parameters
$t_2=0,\,U=2|t_1|,\,\gamma_1=0.12|t_1|,\,\gamma_3=0.1|t_1|,\,\gamma_4=0$,
and $V=G_1=G_3=G_4=0.5|t_1|$. The values of the intralayer and
interlayer hopping integrals used here are close to the values
determined in \cite{Dresselhaus02,Brandt88} for graphite. The
electron density of states for the graphene bilayer for the given
set of parameters is shown in Fig.~\ref{DOS_bilayer}. To
demonstrate the effect of the interlayer Coulomb interaction, we
chose the maximal possible values of intensity $G_1$, $G_3$, and
$G_4$ for which the weak-coupling approximation is still
applicable. Comparison with Fig.~\ref{lambdas_mlg}b shows that the
allowance for the interlayer interactions does not change the
domains of superconductivity with the $f-$ and $d + id$-wave
symmetry types; however, it leads to a significant increase in the
absolute values of the effective coupling constant and, hence, to
an increase in the superconducting transition temperature.

\section{CONCLUSIONS}

We have analyzed the conditions for the emergence of
superconducting Kohn-Luttinger pairing in systems with a linear
dispersion relation using as an example an idealized graphene
monolayer and bilayer, disregarding the Van der Waals potential of
the substrate and impurities. The electronic structure of graphene
is described using the tight binding method in the
Shubin-Vonsovsky model taking into account not only the Coulomb
repulsion of electrons on the same carbon atom, but also the
intersite Coulomb interaction. It is shown that the inclusion of
the Kohn-Luttinger renormalizations up to the second order of
perturbation theory inclusively and the allowance for the
intersite Coulomb interaction determine to a considerable extent
the competition between the superconducting phases with the $f-$
and $d + id$-wave types of the order parameter symmetry. They also
lead to an increase in the absolute values of the effective
interaction and, hence, to higher superconducting transition
temperatures for the idealized graphene monolayer.

The results obtained for the graphene monolayer are generalized to
the case of an idealized graphene bilayer consisting of two
monolayers interacting via Coulomb repulsion between the layers.
It is shown that the analysis of the idealized bilayer system of
graphene leads to a considerably higher value of the
superconducting transition temperature in the context of the
Kohn-Luttinger mechanism.

\section*{ACKNOWLEDGMENTS}

The authors are grateful to V.V. Val'kov for valuable remarks.

This work was performed under the Program of the RAS Division of
Physical Sciences (project no. II.3.1) and was supported
financially by the Russian Foundation for Basic Research (project
nos. 14-02-00058 and 14-02-31237) and by the President of the
Russian Federation (grant no. MK-526.2013.2). One of the authors
(M.M.K.) thanks the "Dynasty" foundation for financial support.


\begin{thebibliography}{99}

\bibitem{Lozovik08}
Yu. E. Lozovik, S. P. Merkulova, and A. A. Sokolik, Phys. Usp.
{\bf 51}, 727 (2008).

\bibitem{Castro09}
A.\,H. Castro Neto, F. Guinea, N.\,M.\,R. Peres, K.\,S. Novoselov,
and A.\,K. Geim, Rev. Mod. Phys. \textbf{81}, 109 (2009).

\bibitem{Kotov12}
V.\,N. Kotov, B. Uchoa, V.\,M. Pereira, F. Guinea, and A.\,H.
Castro Neto, Rev. Mod. Phys. \textbf{84}, 1067 (2012).

\bibitem{Wallace47}
P.\,R. Wallace, Phys. Rev. \textbf{71}, 622 (1947).

\bibitem{Novoselov05}
K.\,S. Novoselov, A.\,K. Geim, S.\,V. Morozov, D. Jiang, M.\,I.
Katsnelson, I.\,V. Grigorieva, S.\,V. Dubonos, and A.\,A. Firsov,
Nature (London) \textbf{438}, 197 (2005).

\bibitem{Tan07}
Y.-W. Tan, Y. Zhang, K. Bolotin, Y. Zhao, S. Adam, E.\,H. Hwang,
S. Das Sarma, H.\,L. Stormer, and P. Kim, Phys. Rev. Lett.
\textbf{99}, 246803 (2007).

\bibitem{Morozov08}
S.\,V. Morozov, K.\,S. Novoselov, M.\,I. Katsnelson, F. Schedin,
D.\,C. Elias, J.\,A. Jaszczak, and A.\,K. Geim, Phys. Rev. Lett.
\textbf{100}, 016602 (2008).

\bibitem{Bolotin08}
K.\,I. Bolotin, K.\,J. Sikes, Z. Jiang, M. Klima, G. Fudenberg, J.
Hone, P. Kim, and H.\,L. Stormer, Solid State Commun.
\textbf{146}, 351 (2008).

\bibitem{Garcia08}
N. Garcia, P. Esquinazi, J. Barzola-Quiquia, B. Ming, and D.
Spoddig, Phys. Rev. B \textbf{78}, 035413 (2008).

\bibitem{Geim06}
A.\,K. Geim, M.\,I. Katsnelson, and K.\,S. Novoselov, Nature Phys.
{\bf 2}, 620 (2006).

\bibitem{Young09}
A.\,F. Young and P. Kim, Nature Phys. {\bf 5}, 222 (2009).

\bibitem{Katsnelson06}
M.\,I. Katsnelson, Eur. Phys. J. B {\bf 51}, 157 (2006).

\bibitem{Rusin09}
T.\,M. Rusin and W. Zawadzki, Phys. Rev. B {\bf 80}, 045416
(2009).

\bibitem{Nair08}
P.\,R. Nair, P. Blake, A.\,N. Grigorenko, K.\,S. Novoselov, T.\,J.
Booth, T. Stauber, N.\,M.\,R. Peres, and A.\,K. Geim, Science {\bf
320}, 1308 (2008).

\bibitem{Munoz12}
W.\,A. Mu$\rm\tilde{n}$oz, L. Covaci, and F.\,M. Peeters, Phys.
Rev. B {\bf 86}, 184505 (2012).

\bibitem{Heersche07}
H.\,B. Heersche, P. Jarillo-Herrero, J.\,B. Oostinga, L.\,M.\,K.
Vandersypen, and A.\,F. Morpurgo, Nature (London) \textbf{446}, 56
(2007).

\bibitem{Shailos07}
A. Shailos, W. Nativel, A. Kasumov, C. Collet, M. Ferrier, S.
Gueron, R. Deblock, and H. Bouchiat, Europhys. Lett. \textbf{79},
57008 (2007).

\bibitem{Du08}
X. Du, I. Skachko, and E.\,Y. Andrei, Phys. Rev. B \textbf{77},
184507 (2008).

\bibitem{Ojeda09}
C. Ojeda-Aristizabal, M. Ferrier, S. Gu\'{e}ron, and H. Bouchiat,
Phys. Rev. B \textbf{79}, 165436 (2009).

\bibitem{Kanda10}
A. Kanda, T. Sato, H. Goto, H. Tomori, S. Takana, Y. Ootuka, and
K. Tsukagoshi, Physica C \textbf{470}, 1477 (2010).

\bibitem{Tomori10}
H. Tomori, A. Kanda, H. Goto, S. Takana, Y. Ootuka, and K.
Tsukagoshi, Physica C \textbf{470}, 1492 (2010).

\bibitem{Peres05}
N.\,M.\,R. Peres, F. Guinea, and A. H. Castro Neto, Phys. Rev. B
\textbf{72}, 174406 (2005).

\bibitem{Marino06}
E.\,C. Marino and L.\,H.\,C.\,M. Nunes, Nucl. Phys. B
\textbf{741}, 404 (2006).

\bibitem{Gonzalez01}
J. Gonz\'{a}lez, F. Guinea, and M.\,A.\,H. Vozmediano, Phys. Rev.
B \textbf{63}, 134421 (2001).

\bibitem{Uchoa07}
B. Uchoa and A.\,H. Castro Neto, Phys. Rev. Lett. \textbf{98},
146801 (2007).

\bibitem{Black07}
A.\,M. Black-Schaffer and S. Doniach, Phys. Rev. B \textbf{75},
134512 (2007).

\bibitem{Honerkamp08}
C. Honerkamp, Phys. Rev. Lett. \textbf{100}, 146404 (2008).

\bibitem{Gonzalez08}
J. Gonz\'{a}lez, Phys. Rev. B \textbf{78}, 205431 (2008).

\bibitem{Markiewicz97}
R.\,S. Markiewicz, J. Phys. Chem. Solids \textbf{58}, 1179 (1997).

\bibitem{Kohn65}
W. Kohn and J.\,M. Luttinger, Phys. Rev. Lett. {\bf 15}, 524
(1965).

\bibitem{Fay68}
D. Fay and A. Layzer, Phys. Rev. Lett. {\bf 20}, 187 (1968).

\bibitem{Kagan88}
M. Yu. Kagan and A. V. Chubukov, JETP Lett. {\bf 47}, 614 (1988).

\bibitem{Baranov92}
M.\,A. Baranov, A.\,V. Chubukov, and M.\,Yu. Kagan, Int. J. Mod.
Phys. B \textbf{6}, 2471 (1992).

\bibitem{Nandkishore12a}
R. Nandkishore, L.\,S. Levitov, and A.\,V. Chubukov, Nature Phys.
\textbf{8}, 158 (2012).

\bibitem{Valenzuela08}
B. Valenzuela and M.\,A.\,H. Vozmediano, New J. Phys. \textbf{10},
113009 (2008).

\bibitem{McChesney10}
J.\,L. McChesney, A. Bostwick, T. Ohta, T. Seyller, K. Horn, J.
Gonz\'{a}lez, and E. Rotenberg, Phys. Rev. Lett. \textbf{104},
136803 (2010).

\bibitem{Kiesel12}
M.\,L. Kiesel, C. Platt, W. Hanke, D.\,A. Abanin, and R. Thomale,
Phys. Rev. B \textbf{86}, 020507(R) (2012).

\bibitem{Wehling11}
T.\,O. Wehling, E. \c{S}a\c{s}{\i}o\u{g}lu, C. Friedrich, A.\,I.
Lichtenstein, M.\,I. Katsnelson, and S. Blugel, Phys. Rev. Lett.
\textbf{106}, 236805 (2011).

\bibitem{Gonzalez13}
J. Gonz\'{a}lez, Phys. Rev. B \textbf{88}, 125434 (2013).

\bibitem{Hwang08}
E.\,H. Hwang and S. Das Sarma, Phys. Rev. Lett. \textbf{101},
156802 (2012).

\bibitem{Migdal58}
A. B. Migdal, Sov. Phys. JETP \textbf{7}, 996 (1958).

\bibitem{Kohn59}
W. Kohn, Phys. Rev. Lett. \textbf{2}, 393 (1959).

\bibitem{Friedel54}
J.\,Friedel, Adv. Phys. {\bf 3}, 446 (1954); Nuovo Cimento Suppl.
{\bf 2}, 287 (1958).

\bibitem{Kagan91}
M.\,Yu. Kagan, Phys. Lett. A \textbf{152}, 303 (1991).

\bibitem{KaganValkov11}
M.\,Yu. Kagan and V.\,V. Val'kov, ÆÝÒÔ {\bf 140}, 179 (2011); Low
Temp. Phys. {\bf 37}, 84 (2011); \emph{A Lifetime in Magnetism and
Superconductivity: A Tribute to Professor David Schoenberg},
Cambridge Scientific Publishers, Cambridge, 2011.

\bibitem{Levin74}
A.\,A. Levin, \emph{Solid State Quantum Chemistry}, McGraw-Hill,
New York, 1977.

\bibitem{Perfetto07}
E. Perfetto, M. Cini, S. Ugenti, P. Castrucci, M. Scarselli, M. De
Crescenzi, F. Rosei, and M. A. El Khakani, Phys. Rev. B
\textbf{76}, 233408 (2007).

\bibitem{Shubin34}
S. Shubin and S. Vonsowsky, Proc. Roy. Soc. A {\bf 145}, 159
(1934); Phys. Zs. UdSSR {\bf 7}, 292 (1935); {\bf 10}, 348 (1936).

\bibitem{Gor'kov61}
L. P. Gor'kov and T. K. Melik-Barkhudarov, Sov. Phys. JETP
\textbf{13}, 1018 (1961).

\bibitem{Scalapino86}
D.\,J. Scalapino, E. Loh, Jr., and J.\,E. Hirsch, Phys. Rev. B
\textbf{34}, 8190 (1986); \textbf{35}, 6694 (1987).

\bibitem{Hlubina99}
R. Hlubina, Phys. Rev. B \textbf{59}, 9600 (1999).

\bibitem{Raghu10}
S. Raghu, S.\,A. Kivelson, and D.\,J. Scalapino, Phys. Rev. B {\bf
81}, 224505 (2010).

\bibitem{Alexandrov11}
A.\,S. Alexandrov and V.\,V. Kabanov, Phys. Rev. Lett. {\bf 106},
136403 (2011).

\bibitem{Kagan13}
M. Yu. Kagan, V. V. Val'kov, V. A. Mitskan, and M. M. Korovushkin,
JETP Lett. \textbf{97}, 226 (2013); JETP \textbf{117}, 728 (2013).

\bibitem{Landau89}
L. D. Landau and E. M. Lifshitz, \emph{Course of Theoretical
Physics, Vol. 3: Quantum Mechanics: Non-Relativistic Theory},
Butterworth-Heinemann, Oxford, 1991.

\bibitem{Dzyaloshinskii88}
I. E. Dzyaloshinskii and V. M. Yakovenko, Sov. Phys. JETP
\textbf{67}, 844 (1988); I. E. Dzyaloshinskii, I. M. Krichever,
and J. Chronek, Sov. Phys. JETP \textbf{67}, 1492 (1988).

\bibitem{Zheleznyak97}
A.\,T. Zheleznyak, V.\,M. Yakovenko, and I.\,E. Dzyaloshinskii,
Phys. Rev. B {\bf 55}, 3200 (1997).

\bibitem{Kagan14}
M.\,Yu. Kagan, V.\,V. Val'kov, V.\,A. Mitskan, and M.\,M.
Korovushkin, Solid State Commun. \textbf{188}, 61 (2014).

\bibitem{McCann06}
E. McCann and V.\,I. Fal'ko, Phys. Rev. Lett. \textbf{96}, 086805
(2006).

\bibitem{McCann13}
E. McCann and M. Koshino, Rep. Prog. Phys. \textbf{76}, 056503
(2013).

\bibitem{Dresselhaus02}
M.\,S. Dresselhaus and G. Dresselhaus, Adv. Phys. {\bf 51}, 1
(2002).

\bibitem{Brandt88}
N.\,B. Brandt, S.\,M. Chudinov, and Y.\,G. Ponomarev, in
\emph{Modern Problems in Condensed Matter Sciences}, edited by
V.\,M. Agranovich and A.\,A. Maradudin, North-Holland, Amsterdam,
Vol 20.1, 1988.


\end{thebibliography}
\end{document}